# Voice Mapping of Text-to-Speech Systems: A Metric-Based Approach for Voice Quality Assessment

Huanchen Cai[1,a], Sten Ternström[1]

[1] *Division of Speech, Music and Hearing, School of Electrical Engineering and Computer Science, KTH Royal Institute of Technology, SE-100 44 Stockholm, Sweden*

*huanchen@kth.se*, *stern@kth.se*

This study investigates voice mapping as an evaluation framework for text-to-speech (TTS) synthesis quality. The study analyzes six TTS models, including historical and recent ones. The metrics are crest factor, spectrum balance, and cepstral peak prominence (CPPs). We investigated 6 influential TTS models: Merlin, Tacotron 2, Transformer TTS, FastSpeech 2, Glow-TTS, and VITS. The results demonstrate that voice range serves as a primary indicator of model capability, with VITS showing the largest range among tested models. Glow-TTS exhibited superior performance in soft phonation, indicated by higher spectrum balance, despite limited voice range. The results showed that the CPPs values between 7-8 dB indicate natural voice quality, while with CPPs exceeding 10 dB, the speech tends to sound robotic. These findings underscore the need for voice mapping to evaluate vocal effort, and capture how TTS systems handle voice dynamic and expressiveness.

[a] Email: huanchen@kth.se

## 1. Introduction

### *1.1 The Challenge of TTS Evaluation*

The evaluation of Text-to-Speech (TTS) systems remains a challenge in speech synthesis research, despite significant advances in generation quality. Current assessment methods broadly fall into two categories: subjective[1] and objective methods[2], each with distinct advantages and limitations.

Subjective evaluation remains the gold standard in TTS assessment. It relies primarily on human assessment protocols such as those established by the Blizzard Challenges[1]. These protocols examine multiple dimensions of synthetic speech, including naturalness, intelligibility, and prosodic elements such as stress patterns, emotional expression, and overall listening effort.

The Mean Opinion Score described by ITU-T P.800[3] (MOS) serves as the primary metric in subjective evaluation and was the only method used in previous Blizzard Challenges. However, recent studies have revealed significant limitations in MOS-based assessment [4]. For example, listeners tend to utilize the entire range of scoring options, introducing range-equalizing bias.[5] Replication studies have demonstrated that MOS produces relative rather than absolute scores[4]. Furthermore, inconsistent methodology across studies, with varying scale labels, increments, and participant instructions, makes results difficult to compare[6]. MOS score from different TTS models presents score instability. For instance, MOS scores reported for Tacotron 2 ranged from 3.70[8] to 4.53[7, 8], while FastSpeech 2 showed similar variability (3.83[8] to 4.32)[7, 9].

Beyond score instability, subjective evaluations face practical challenges: they are resource-intensive, time-consuming, and subject to listener variability[10]. These limitations have motivated research into non-intrusive automated MOS prediction systems[11-14], MOS prediction challenges[15], and innovations in experimental paradigms[16].

Objective evaluation methods provide quantitative assessment through measurable acoustic features. Commonly used objective metrics include Mel-Cepstral Distortion (MCD) for spectral fidelity[17], Extended Short-Time Objective Intelligibility (ESTOI) for intelligibility[18], Perceptual Evaluation of Speech Quality (PESQ) for transmission quality and distortions[19], and various error measurements for voice characteristics (band aperiodicities (BAP)[20], voiced/unvoiced error (V/UV)[21]). However, most existing objective metrics operate at the frame or utterance level, typically over short audio segments lasting only a few seconds and then summarize the overall performance by averaging these segment-level scores. As such, they do not directly reflect changes in voice quality, which is closely tied to physiological and phonatory factors like glottal periodicity, vocal fold tension, and airflow regulation. These dimensions are essential in characterizing the effort and expressiveness of the voice. Standard objective metrics overlook the within-speaker variability across different pitch and loudness conditions. Consequently, they fail to capture temporal voice quality dynamics that may emerge during long-form speech synthesis, such as degradation, instability, or inconsistent vocal effort.

While contemporary TTS systems can now produce speech that sounds convincingly human, an important yet underexplored aspect is their capacity to reproduce dynamic vocal characteristics, such as varying vocal effort across different phonation modes (e.g., soft, loud, modal, falsetto). These modes reflect the natural correlation between phonatory effort and perceived voice quality, and therefore affect the overall expressiveness[22]. In many current TTS models, control over pitch and loudness is often implemented independently such as simply raising the fundamental frequency ($f_o$) or output volume. However, when a human speaker increases vocal intensity, the change does not merely involve louder output, but complex physiological adjustments and the acoustic modifications that affect the entire voice quality profile[23-25]. Unlike those conventional objective metrics, our proposed approach allows for fine-grained analysis of a model's voice range for long-form content and its ability to capture the natural coupling between vocal effort and voice quality. Therefor not individual sentences or clips, but voices that are of good quality to listen for several minutes (in this research 100 sentences).

### 1.2 *Voice Mapping: Application to Synthesized Voices*

To address these evaluation challenges, we propose voice mapping[26] as an objective metric-based assessment tool for synthetic voices. This approach offers visualization and comparison scheme on both phonetic and perceptual characteristics of TTS output, with particular emphasis on voice dynamics and quality. The novelty of this visualization method is in applying voice mapping to synthesized rather than natural voices. The purpose of doing that is to compare the variability in synthesized voices to that in natural voices.

Inspired by the Voice Range Profile[27] (or phonetogram[28]), voice mapping has demonstrated its value in voice research, particularly for studying voice production[29, 30] and monitoring voice disorders[31], and is proving to be especially useful for pre- and post-intervention comparisons[32]. Figure 1 illustrates this through a case study of a patient[31], who went through a thyroidectomy (removal of the thyroid gland).

Voice maps in this study are two-dimensional plots representing voice samples distributed across pitch and intensity, where each cell indicates the average voice quality within that region. The voice quality is quantified using smoothed Cepstral Peak Prominence (CPPs, a metric measuring the degree of harmonic organization and periodicity in the voice

signal[33]) values, with warmer colors (e.g., red) indicating higher CPPs (better voice quality), and cooler colors (e.g., blue) indicating lower CPPs (poorer voice quality). The third panel shows the difference between pre- and post-condition maps, highlighting areas where voice quality decreased (in red) or increased (in blue) after intervention. The voice map example demonstrates a significant reduction in voice range, shown by the decreased area of colored cells, alongside diminished voice quality indicated by lower CPPs. This is an objective and DSP way of assessing voice quality. In this case, the pre-condition voice quality is better than post-condition, in terms of the harmonic periodicities.

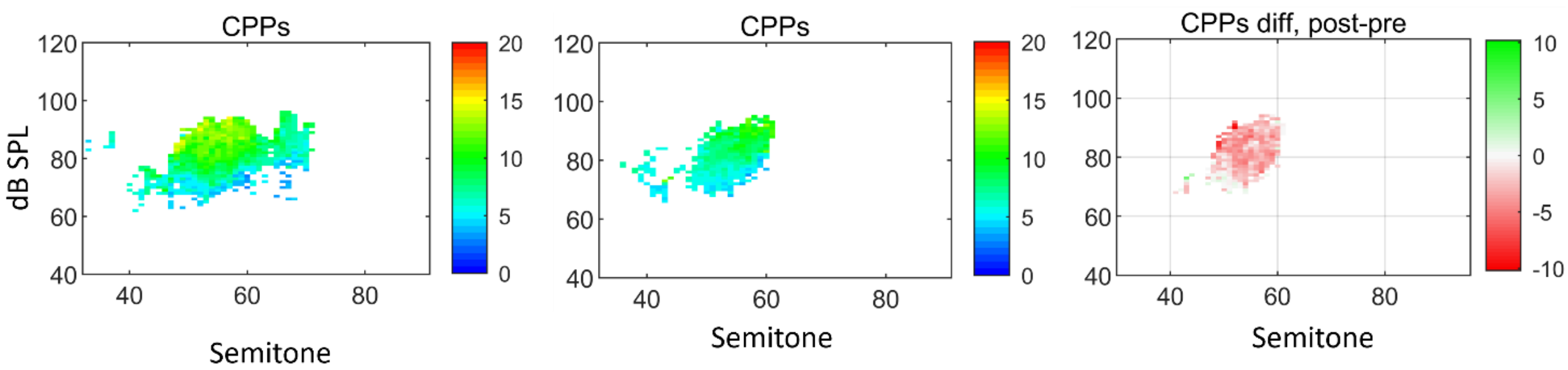

Figure 1. (color online) Voice map example showing voice range and CPPs difference (in dB) pre-thyroidectomy (left), post-thyroidectomy (middle), and their difference in overlap (right). The horizontal axis represents pitch in semitones, referenced to 55 Hz (i.e., semitone 0 = 55 Hz; semitone 12 = 110 Hz, and so on). The vertical axis shows sound pressure level (SPL) in decibels (dB). The color scale indicates CPPs values in dB, a proxy for harmonic organization and voice quality — higher values (warmer colors) reflect better periodicity and clearer voicing. In the difference map (right), red areas indicate regions where CPPs decreased after surgery, suggesting degraded voice quality.

Just as voice mapping effectively captures changes in human voice characteristics after medical interventions, it can provide similar insights when analyzing synthetic voices. In this context, we can consider TTS architectures as analogous to "interventions" in the voice production process.

### *1.3 Voice Mapping for Modern TTS Analysis*

Early TTS models, such as rule-based [34-37] models in 1970s-1980s and concatenative synthesis[38-40] methods in 1990s-2000s, relied primarily on predefined linguistic rules and concatenated segments of recorded speech. While rule-based systems often produced robotic-sounding output, concatenative synthesis could achieve natural speech but with significant concatenation artifacts. Statistical parametric synthesis[41-43] invented in 2000s-early 2010s offer greater flexibility but still lacked naturalness. These early systems exhibited clear qualitative limitations, including word-level errors, poor signal quality, and noise artifacts among others[44].

A shift in speech quality occurred with the introduction of neural network architectures, beginning with Heiga Zen's pioneering work[45] in 2013 and the Merlin system[46] in 2016. WaveNet[47], introduced in 2016, demonstrated unprecedented audio quality through modeling of raw waveforms. Tacotron[48] (2017) enabled end-to-end speech generation and Transformer-based models[49] (2019) further improved naturalness and sequence modeling capabilities. FastPitch[50] (2021), VITS[51] (2021) advanced the field by improving pitch controllability and enabling fully end-to-end modeling with variational inference. Despite many more recent models such as Bark speecht5[52], NaturalSpeech[53, 54], StyleTTS, F5-TTS[55], Parler-TTS[56] being proposed at an accelerating pace, this paper does not aim to track the development of all these systems. Instead, it focuses on evaluating the interplay between voice range and phonatory effort, which remains an underexplored dimension of TTS evaluation. For a more detailed review of TTS models, readers are encouraged to refer to review papers[44].

This represents only a brief overview of the development of TTS models. However, this evolution clearly demonstrates the field's dramatic progress in voice quality. To visualize this advancement, we propose the voice maps. Our key aspects for evaluation are the voice capabilities (voice range) and a few selected voice quality metrics. To analyze how well a TTS system captures these features in system level, we compare the distribution of the metrics, and perform difference mapping between the original speaker's recordings (baseline) and the synthesized speech, particularly analyzing overlapping vocal regions.

## 2. Method

### *2.1 Creation of Synthesis*

To limit the computational task while ensuring meaningful results, we analyzed a representative subset of the LJSpeech dataset[57] rather than the complete corpus. The LJSpeech dataset, consisting of recordings by Linda Johnson and processed by Keith Ito, served dual purposes: as ground-truth reference audio and as the baseline condition for model comparison. Each audio file is a single-channel 16-bit PCM WAV with a sample rate of 22050 Hz. By using this publicly available English speech dataset across all models, we ensured a controlled environment for evaluation.

We selected 100 consecutive sentences (LJ050-0029 to LJ050-0130), totalling approximately 27 minutes of audio. The number of sentences sufficient for comparison is discussed in section 3.1. We then generated their synthesized versions using the TTS models mentioned above, all pre-trained on the LJSpeech corpus. While this setup does not reflect generalization to unseen inputs, it is intentional, as our focus is on analyzing model behavior under controlled conditions. For models supporting multiple vocoder configurations, we produced separate outputs for each configuration to evaluate their impact on voice quality.

Future work may include testing on unseen texts to assess the generalization of voice quality preservation.

### 2.2 *TTS Model selection*

As this is exploratory research, we did not exhaustively explore all the possible models, but selected six architectures (Merlin, Tacotron 2, Transformer, FastSpeech 2, Glow TTS, and VITS) that represent key developments in TTS technology during the last decade. All models were implemented using publicly available pre-trained versions via Coqui TTS and ESPnet[58]. They are two widely adopted platforms that offer well-documented, community-validated implementations. These toolkits provide stable and reproducible pipelines, enabling fair comparisons without the confounds of custom re-training or hyperparameter tuning.

Models were implemented using their publicly available pre-trained weights. The models in discussion are listed in Table 1.

Table 1. TTS model list.

| Model | Year | Key Feature(s) |
| --- | --- | --- |
| Merlin | 2016 | Statistical (DNN-HMM hybrid) |
| Tacotron 2 | 2018 | Autoregressive sequence-to-sequence |
| Transformer TTS | 2019 | Autoregressive with self-attention |
| FastSpeech 2 | 2020 | Non-autoregressive feed-forward |
| Glow-TTS | 2020 | Non-autoregressive, flow-based |
| VITS | 2021 | End-to-end, variational + adversarial |

Our analysis spans key developments in neural TTS architectures over the past decade: Merlin (2016), which integrated neural networks with HMM-based synthesis; Tacotron 2 (2018), which introduced end-to-end text-to-spectrogram synthesis through an encoder-decoder framework; Transformer-TTS (2019), which replaced recurrent networks with self-attention mechanisms; FastSpeech 2 (2020), which developed non-autoregressive generation with explicit acoustic feature prediction; Glow-TTS (2020), which employed normalizing flows for parallel generation; and VITS (2021), which unified acoustic modeling and vocoding in a single network through conditional variational autoencoders.

We distinguish between the acoustic model, which transforms text into intermediate acoustic features (e.g., Mel-spectrograms or latent representations), and the vocoder, which converts these features into waveform audio. To isolate the effect of vocoders on voice quality, we selected Tacotron 2 as a reference acoustic model, we compared two state-of-the-art vocoders: Multiband-MelGAN[59], UnivNet[60]. Multiband-MelGAN utilizes a parallel processing strategy, decomposing the audio spectrum into multiple frequency bands for simultaneous synthesis, while UnivNet implements a full-band processing approach enhanced by adversarial training techniques.

### *2.3 Data process*

Typically, when we use voice mapping, we would require a SPL-calibrated dataset, as the method involves analyzing voice quality in the ($f_o$, SPL) space. In such a framework, comparing two utterances before and after modelling relies on identifying overlapping regions in this joint space. Thus, it is important to establish a reference SPL framework from the outset, ensuring comparability across datasets and models. However, in a TTS voice setting, the sound level calibration information is lost. This would make the comparison between two utterances difficult because the SPL would also influence the voice quality metrics[61]. Fortunately, since most systems do not condition on volume during training, we assume that the output can be considered as self-calibrated by the system.

The synthesized outputs are generated at a sample rate of 44.1 kHz and named by model and text index. We record voice map files and cycle boundaries using FonaDyn 3.1.0[62]. FonaDyn extracts voice metrics such as $f_o$, SPL, cepstral peak prominence (CPPs), crest factor, and spectrum balance (SB). Crest factor and SPL are computed cycle-synchronously. Others are windowed with a fixed frame length according to FonaDyn handbook: CPPs (23 ms), SB (4th-order low-pass smoothing at 50 Hz), and $f_o$ (23 ms).

The cycle detection[63] begins with a leaky integrator processing the pre-conditioned EGG signal. The integration follows $xn = yn + \alpha xn - 1$, where $yn$ is our input EGG signal, $xn - 1$ is the previous integrated value, and $\alpha = 0.999$ gives us a long but not infinite memory at 44.1 kHz sampling rate. A second-order Butterworth high-pass filter at 50 Hz prevents signal drift in the integrator.

These metrics are the acoustic metrics currently supported by FonaDyn for detailed voice quality analysis, but it is possible to include more metrics into its framework. Voice maps are stored as a CSV text file with the map matrix, where each row is an occupied cell in the map, and each columns contains the averages of one metric.

The crest factor is computed as the ratio of the peak amplitude to the RMS amplitude for each phonatory cycle. It is computed as:

$$\mathrm{CF}_i = \frac{max_t|x_i(t)|}{\sqrt{\frac{1}{T}\int_{t_i}^{t_i+T} x_i(t)^2\ dt}}$$

where $\mathrm{CF}_i$ is the crest factor for the $i$-th phonatory cycle, $x_i(t)$ is the speech signal within this cycle, $t_i$ is the starting time of the cycle, and $T$ is its duration. $max_t|x_i(t)|$ denotes the peak (maximum absolute) amplitude, and the denominator represents the root mean square (RMS) amplitude over the cycle. It is an indicator of the "peakiness" of the voice signal and tends to be especially high in creaky voice. The crest factor is similar to Maximum Flow Declination Rate (MFDR) but can be directly computed from the acoustic signal without applying glottal inverse filtering. In the vocal range of habitual speech, a low crest factor suggests gradual interruptions in glottal flow during closure, which can correspond to a voice perceived as less clear or of reduced quality.

Spectrum Balance is defined as the difference in acoustic power level between frequencies above 2 kHz and those below 1.5 kHz, expressed as:

$$SB = 10 \cdot log_{10}\left(\frac{W_{>2kHz}}{W_{<1.5kHz}}\right)(dB)$$

The acoustic power in each band is computed by summing the squared magnitudes of the STFT components within the respective frequency range. A very low (i.e., highly negative) SB indicates reduced energy in the higher frequencies relative to the lower, which can result in a voice that is perceived as dull or less clear.

CPPs quantifies periodicity in the spectrum, with higher CPPs indicating stronger harmonic structure and reduced noise. In this study, CPPs was calculated following the method of Awan et al[64]. The cepstrum bin count was 512. There was no interpolation between cepstrum peaks. The raw CPP is calculated as the difference between the cepstral peak and the baseline cepstral value representing noise or non-periodic components. The raw

CPP was smoothed using an exponential decay time window corresponding to a 16-Hz first-order low-pass filter per bin, and averaging across 7 bins in quefrency. Lower CPPs values suggest a noisier signal with weaker periodicity, potentially leading to a voice perceived as less stable or clear.

To visualize and analyze the variation of voice quality metrics across different phonatory conditions, the voice map is constructed by mapping the acoustic features onto a two-dimensional space defined by $f_o$ and SPL. For each glottal cycle $i$, we extract the $f_o$, the SPL, and the corresponding acoustic metric value $M(i)$ (e.g., Crest Factor, CPP, RMS). The voice map is then represented as the set of triplets:

$$\text{VoiceMap} = \{ (F_0(i),\ SPL(i),\ M(i)) \mid i = 1,2,\dots,N \}$$

To facilitate analysis and visualization, the ($f_o$, SPL) space can be discretized into bins of size $(\Delta f)$ and $(\Delta s)$, respectively. Within each bin centered at $(f, s)$, the metric values are averaged over all cycles $i$ whose $f_o$ and SPL fall within the bin:

$$M_{\text{avg}}(f, s) = \frac{1}{|\mathcal{I}_{f,s}|} \sum_{i \in \mathcal{I}_{f,s}} M(i)$$

where $\mathcal{I}_{f,s} = i \mid F_0(i) \in [f, f + \Delta f),\ SPL(i) \in [s, s + \Delta s)$

With the voice map files generated by FonaDyn, we used Python scripts to plot the voice maps for each model, along with difference maps comparing each model to the original speaker. Statistical calculations are then performed on the resulting voice maps.

### *2.4 Evaluation Design*

TTS is a one-to-many problem, where a single text input can produce multiple realizations. Consequently, when evaluating the voice quality of synthesized speech, it is essential that the generated output demonstrates a voice range that is at least comparable to, if not broader than, that of the original (ground-truth) speech. Thus, voice range is an important indicator of a model's expressiveness.

Voice quality metrics provide analysis that better aligns with how humans perceive and evaluate synthetic voices. In addition to graphical representation as voice mapping typically does, we also perform statistical analyses to distill the performance into numerical values, in order to offer a quantitative summary of each model's capabilities within our protocol.

## 3. Result

### *3.1 SAM: an Early Example*

To illustrate the historical contrast and demonstrate the interpretability of our proposed method, Figure 2 provides a striking visualization of this evolution by comparing voice maps of SAM synthesizer[65] in 1982 against VITS model in 2021. While SAM is not included in the main comparative analysis due to its outdated architecture and lack of compatibility with current evaluation pipelines, we believe this example serves to highlight the utility of voice mapping in visualizing qualitative differences.

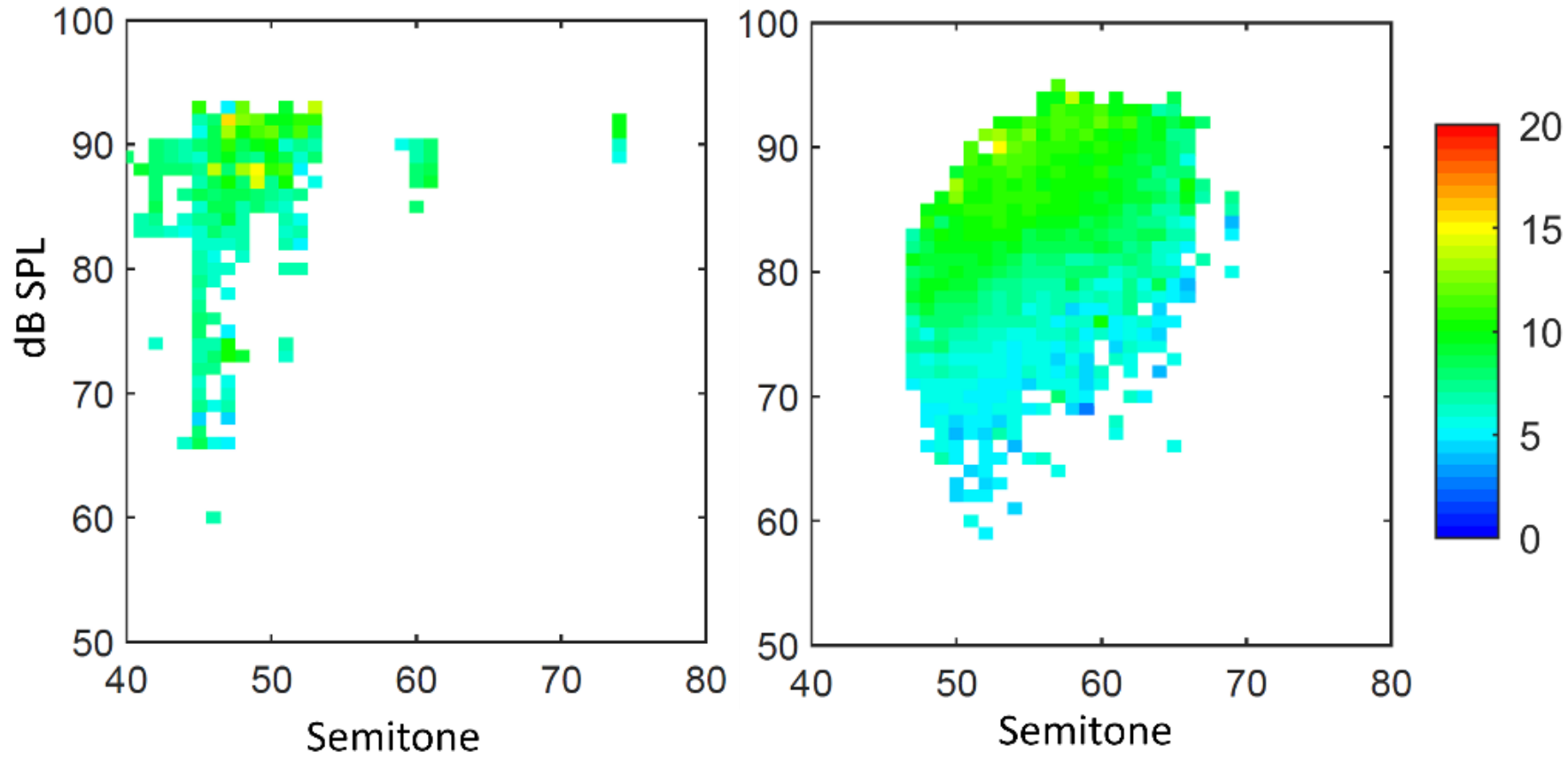

Figure 2. (color online) Voice maps of the SAM synthesizer (left) and the VITS synthesizer (right) showing the metric CPPs. Both maps reflect the same input texts. The horizontal axis is $f_o$ in semitones; and the vertical axis is SPL in dB. The color scale is CPPs (dB).

The voice maps, both generated from identical input texts, reveal apparent differences in capability: the SAM synthesizer exhibits a limited, scattered voice range and irregular patterns, while VITS demonstrates significantly broader vocal capabilities. Such scatter in SAM often results from failures in pitch tracking algorithms, such as autocorrelation. Additionally, as introduced earlier, CPPs quantifies the harmonicity of the signal. The output from SAM synthesizer is characterized by highly dispersed CPPs values, suggesting unstable or degraded harmonicity, which contributes to robotic or unnatural-sounding speech.

### *3.2 Voice Range*

While our sentence subset may not demonstrate the absolute maximum voice range capabilities of each model and the speaker, we observed that the cumulative voice range area (the amount of $f_o$ – SPL pairs) typically plateaued after approximately 20 sentences, suggesting that our sample adequately represented the voice variation found in natural speech.

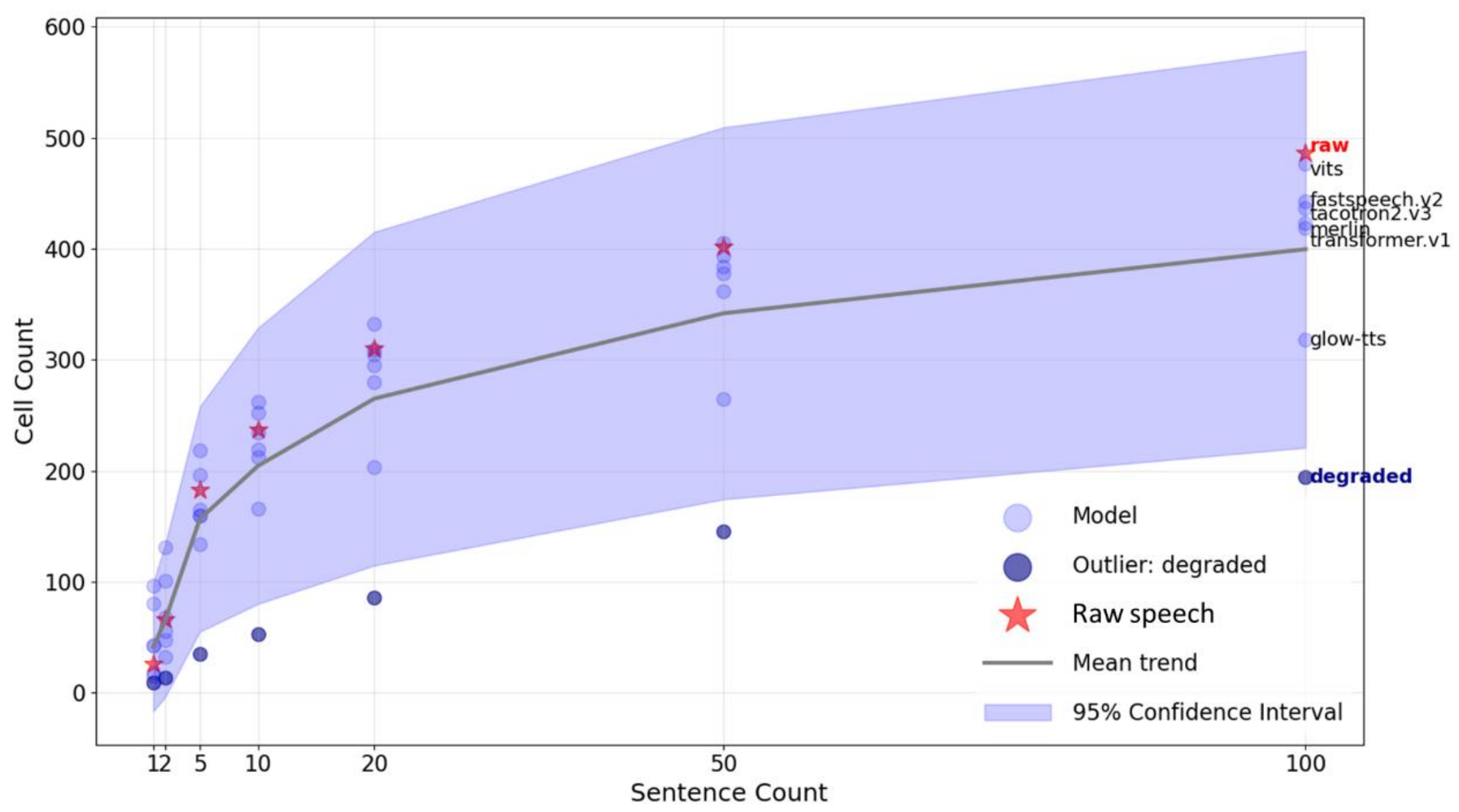

Figure 3. Change of $f_o$ and SPL cell voice map coverage area (measured in cells of 1 semitone × 1 dB) versus increasing sentence counts. Red stars denote natural (raw) speech. The grey line shows the mean trend, with a 95% confidence interval in blue. Degraded outlier cases, labeled as "degraded," exhibit significantly reduced coverage due to vocoder noise or failed $f_o$ tracking.

Figure 3 shows that the raw voice maintains the widest range, with the modern VITS model closely following. Other transformer-based models position near the middle of the distribution, while Glow-TTS displays a limited voice range. The starred points indicate instances of poor-quality synthesis due to model and vocoder mismatch, where too much noise and inadequate $f_o$ extraction results in a significantly reduced voice range.

### *3.3 Statistical results*

We present the results for crest factor, spectrum balance, and CPPs below.

Table 2. Crest Factor comparison across different TTS models and the LJSpeech dataset (baseline). The degraded synthesis represents a failed synthesis for reference, which exhibits noisy, robotic qualities and is barely intelligible.

| Model | Mean±Std.dev. | CI Range (95%) | Diff from Raw |
|---|---|---|---|
| Merlin | 2.58±0.40 | [2.54, 2.61] | +0.04 |
| Tacotron 2 | 2.56±0.42 | [2.52, 2.60] | +0.03 |
| Transformer | 2.53±0.44 | [2.48, 2.57] | -0.00 |
| Fast speech 2 | 2.57±0.43 | [2.53, 2.60] | +0.03 |
| Glow TTS | 2.63±0.32 | [2.59, 2.66] | +0.10 |
| VITS | 2.54±0.47 | [2.50, 2.59] | +0.01 |
| LJSpeech | 2.53±0.42 | [2.49, 2.57] | baseline |
| Degraded synthesis | 2.18±0.44 | [2.12, 2.24] | -0.35 |

As shown in Table 2, Glow-TTS exhibits a notably higher crest factor (2.63 ± 0.32), indicating enhanced "peakiness" in voice signal characteristics, often associated with clear voice quality. Models such as Tacotron 2 (2.56 ± 0.42) and FastSpeech 2 (2.57 ± 0.43) maintain moderate levels similar to natural voices.

Table 3. Spectrum Balance (in dB) comparison across different TTS models and the LJSpeech dataset (baseline).

| Model | Mean±Std.dev | CI Range (95%) | Difference from Raw |
|---|---|---|---|
| Merlin | -19.75±7.25 | [-20.44, -19.06] | -4.29 |
| Tacotron 2 | -16.88±8.70 | [-17.69, -16.06] | -1.41 |
| Transformer | -17.79±9.06 | [-18.66, -16.92] | -2.32 |
| Fast speech 2 | -16.16±8.22 | [-16.93, -15.40] | -0.69 |
| Glow TTS | -14.89±7.12 | [-15.68, -14.11] | +0.57 |
| VITS | -16.90±9.53 | [-17.76, -16.05] | -1.44 |
| LJSpeech | -15.47±12.50 | [-16.58, -14.36] | baseline |
| Degraded synthesis | -31.32±14.73 | [-33.39, -29.25] | -15.85 |

As shown in Table 3, Glow-TTS displays a higher spectrum balance (-14.89±7.12) in the lower frequency energy, potentially reflecting an enhanced ability to capture soft phonation characteristics. In contrast, Merlin (-19.75±7.25) and low-quality synthesis (-31.32 ± 14.73) outputs have significant energy losses in high frequencies, resulting in dullness in voice quality.

Table 4. CPPs comparison across different TTS models and the LJSpeech dataset (baseline).

| Model | Mean±Std.dev. | CI Range (95%) | Difference from Raw |
|---|---|---|---|
| Merlin | 12.24±1.92 | [12.05, 12.42] | +4.50 |
| Tacotron 2 | 7.56±2.20 | [7.35, 7.76] | -0.18 |
| Transformer | 7.34±2.09 | [7.14, 7.54] | -0.40 |
| Fast speech 2 | 6.68±1.92 | [6.50, 6.86] | -1.06 |
| Glow TTS | 7.96±2.38 | [7.70, 8.22] | +0.23 |
| VITS | 7.75±2.24 | [7.55, 7.95] | +0.02 |
| LJSpeech | 7.73±2.35 | [7.53, 7.94] | baseline |
| Degraded synthesis | 9.73±4.45 | [9.11, 10.36] | +2.00 |

The elevated CPPS values in the Merlin model (12.24 ± 1.92) actually exceed typical human ranges, suggesting excessive periodicity that may contribute to be perceiving as robotic in the synthesized speech. Table 4 shows that the newer neural TTS models' CPPS values (ranging from 6.68 to 7.96) fall squarely within expected ranges for natural female speech, indicating that these architectures may be better optimized for generating speech with human-like periodic patterns. FastSpeech 2 showed the lowest CPPS values (6.68 ± 1.92) among the tested models, while Glow TTS demonstrated the highest values (7.96 ± 2.38) among the contemporary architectures, though all remained within natural ranges.

### *3.4 Voice Map Comparisons*

Voice mapping provides a visual means to analyze synthetic speech characteristics without actually hearing it.

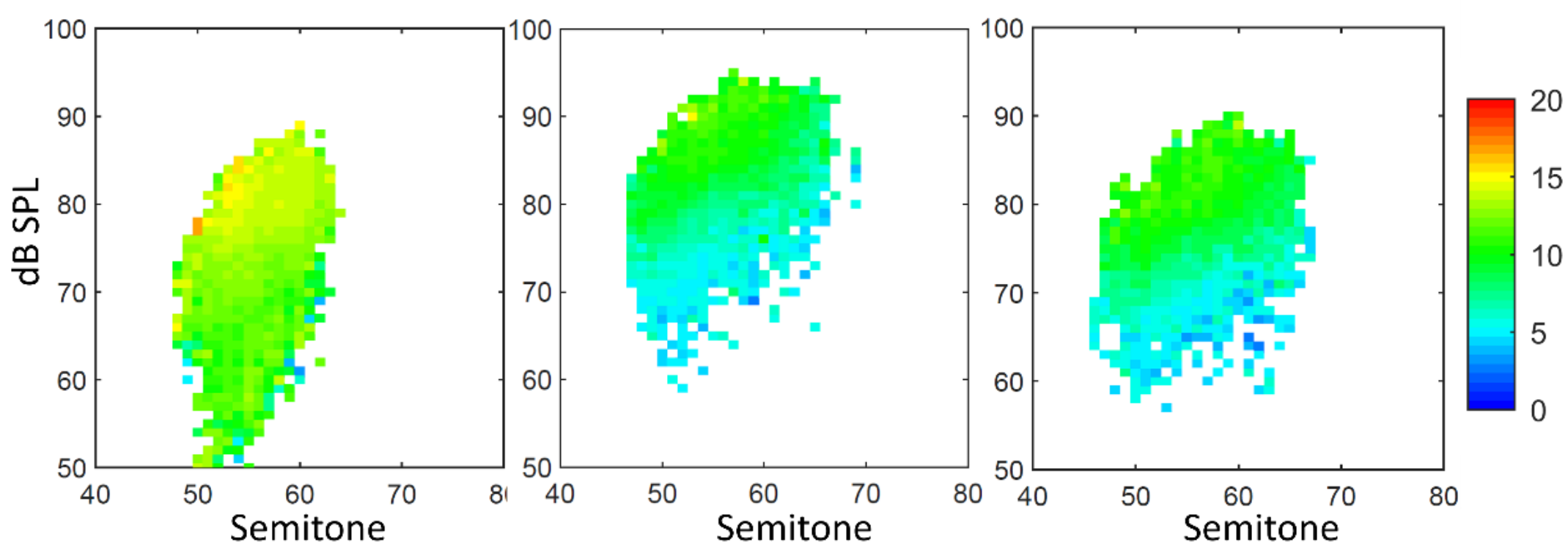

Figure 4. Voice maps of CPPs as measured from the output of several TTS models. Left: Merlin; Middle: VITS; Right: LJSpeech speaker.

As shown in Figure 4, the CPPs voice maps reveal distinct voice range and voice quality patterns across different TTS systems. The Merlin system exhibits a notably constrained voice range and relatively uniform CPPs values across its phonation range, suggesting limited dynamic variation. In contrast, VITS demonstrates a voice map that closely mirrors the original LJSpeech speaker, with both the range coverage and CPPs distribution patterns showing remarkable similarity. This visual analysis corroborates the subjective assessment that modern neural TTS systems like VITS can achieve very natural voice reproduction.

Voice mapping can reveal model-specific differences that are not captured by isolated values of a metric. Higher CPPs generally indicates a clearer, more resonant voice quality.

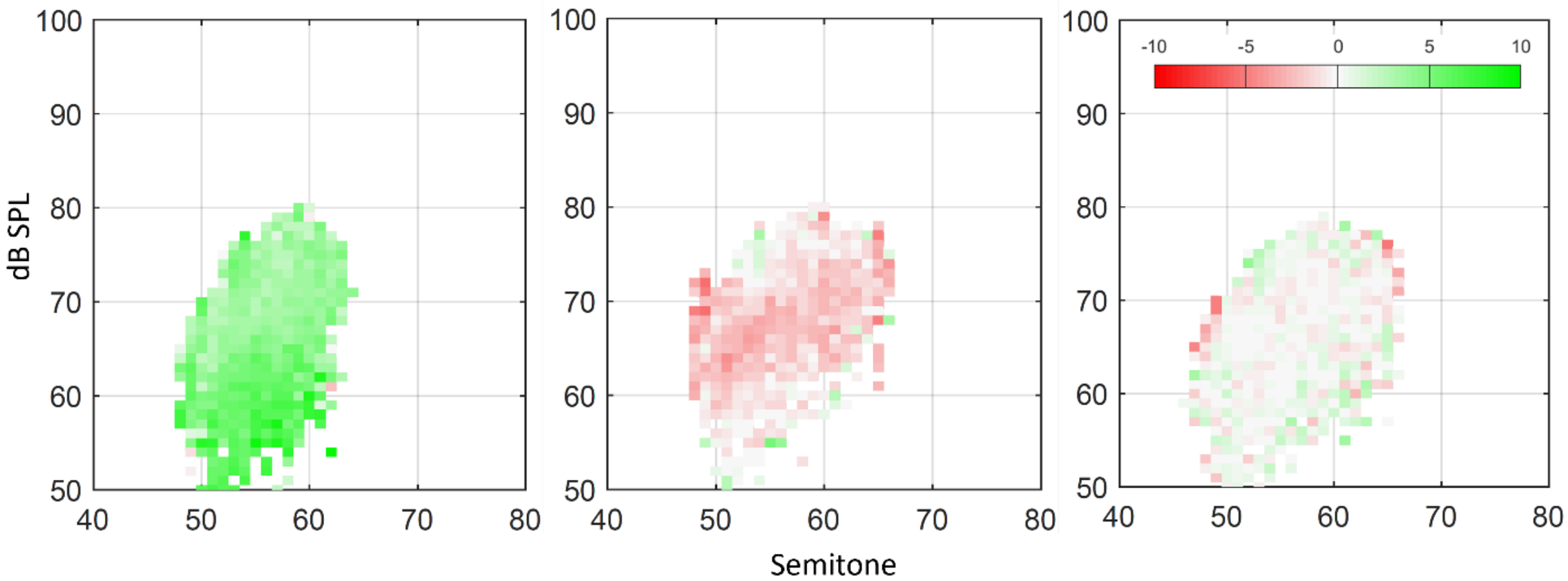

Figure 5. Difference maps of the overlapping area of CPPs between synthetic models and original LJSpeech speaker (Left: Merlin; Middle: Tacotron 2; Right: VITS). The green color displays where model CPPs values exceed those of raw speech, while red color shows where they are lower. The saturation of each color indicates the magnitude of the absolute difference, with more saturated colors representing larger differences. The color scale at top right signifies CPPs difference in dB.

In this comparison shown in Figure 5, Merlin's CPPs are notably higher than typical human ranges, which may contribute to a robotic tonal quality. Tacotron 2's CPPs are closer to natural levels, while VITS aligns even more closely with the original voice's characteristics, suggesting its effectiveness in retaining human-like resonance in synthesized speech.

Another example in Figure 6 shows that, Glow TTS, with a higher SB in lower frequency range, indicates a potential advantage in preserving harmonic structure and reducing breathiness or noise when synthesizing quiet or low-effort speech. This might mean that Glow-TTS can represent subtle voice qualities in quiet or soft speech more naturally, though its limited voice range could reduce expressiveness.

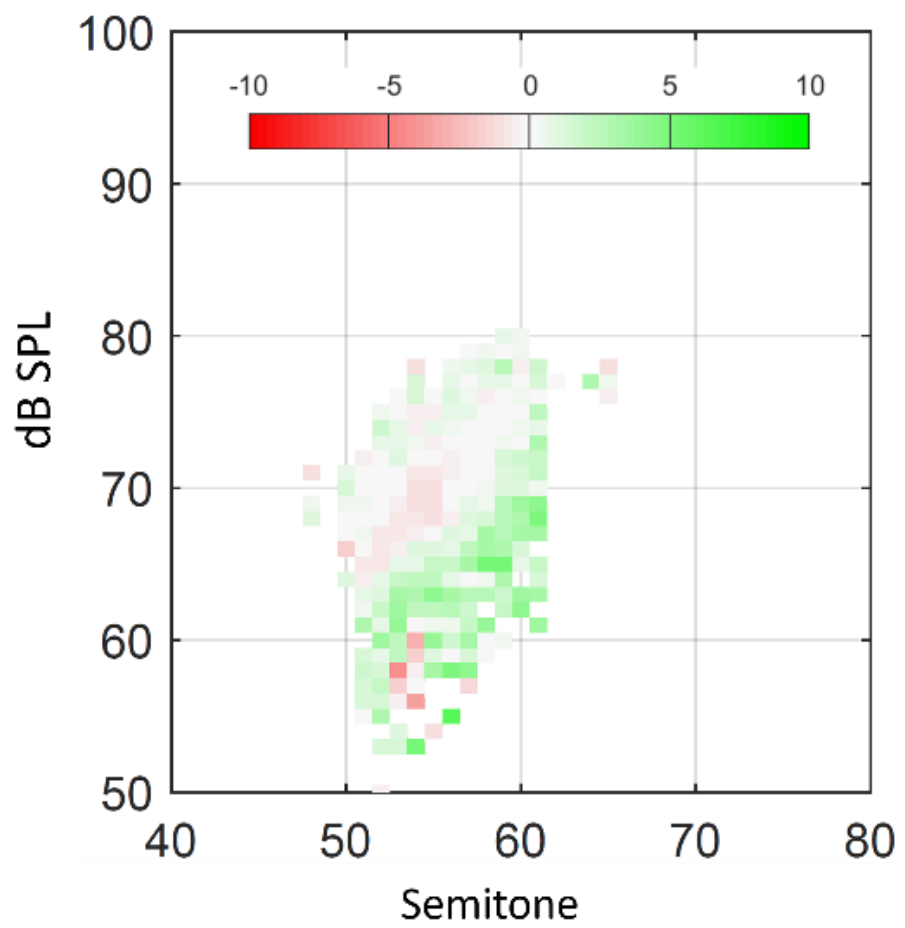

Figure 6. Difference map of Spectrum Balance between Glow-TTS synthesis and raw LJSpeech speaker. The increase in Spectrum Balance of Glow-TTS is primarily concentrated in the softer SPL regions (green coloring indicates where Glow-TTS values exceed raw speech).

### *3.5 Vocoder Comparison*

Table 5. Spectrum Balance (in dB) comparison between vocoders.

| Model | Mean±Std.dev. | CI Range (95%) | Difference from Raw |
|---|---|---|---|
| Multiband-MelGAN | -16.80±8.10 | [-17.58, -16.03] | -1.27 |
| UnivNet | -15.92±8.71 | [-16.78, -15.05] | -0.39 |

Table 6. CPPs comparison between vocoders

| Model | Mean±Std.dev | CI Range (95%) | Difference from Raw |
|---|---|---|---|
| Multiband-MelGAN | 7.15±1.58 | [7.00, 7.30] | -0.60 |
| UnivNet | 7.92±1.77 | [7.74, 8.10] | +0.17 |

From Table 5 and Table 6 it can be seen that UnivNet performs better with a mean of -15.92dB in Spectrum Balance, and a mean of 7.92 in CPPs which is slightly positive deviation from raw (+0.17). By definition of those metrics, it indicates that UnivNet better preserves the spectral characteristics of the original signal (Spectrum Balance), and periodicity (CPPs) than Multiband-MelGAN.

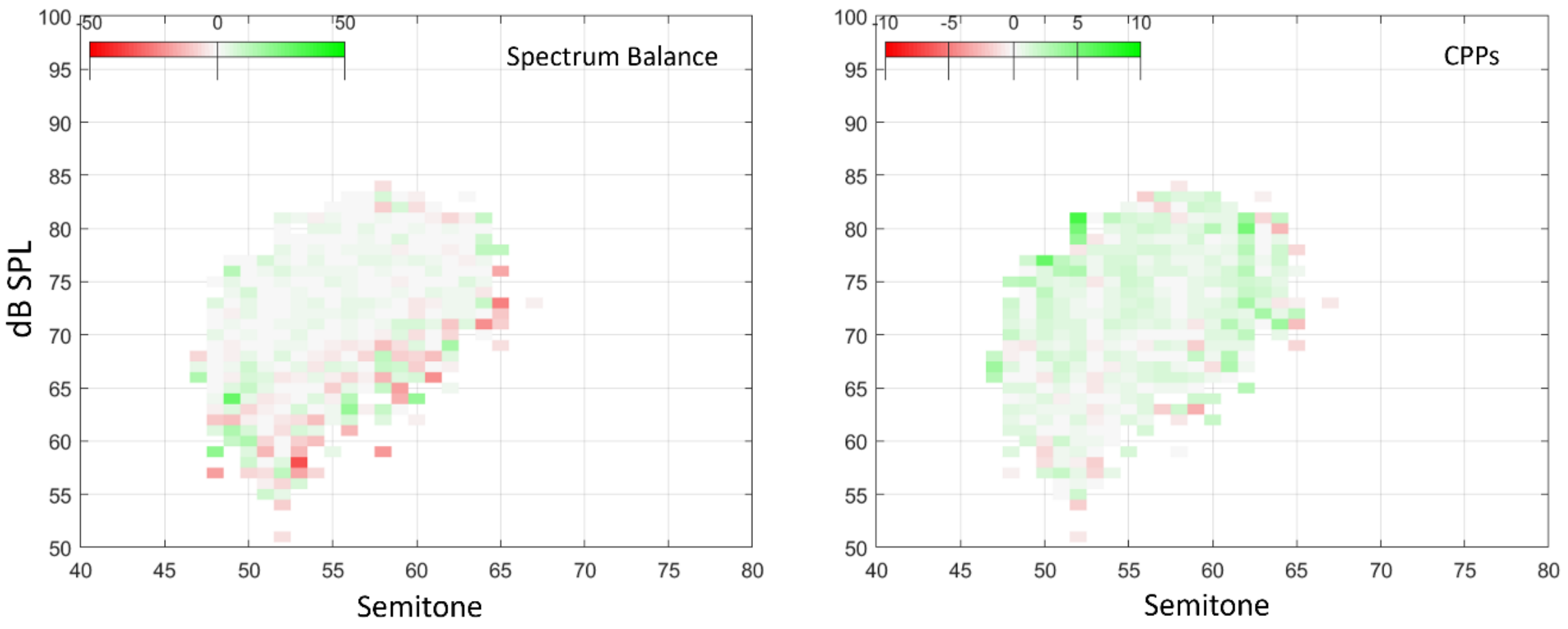

Figure 7. Difference maps comparing UnivNet and Multiband-MelGAN performance in (a) Spectrum Balance (dB) and (b) CPPs. Green cells indicate where UnivNet values exceed Multiband-MelGAN, while red cells show where Multiband-MelGAN values are higher.

Figure 7 further shows illustrates the distributional characteristics of these differences. In terms of Spectrum Balance, UnivNet shows slightly higher values overall (indicated by a subtle green predominance), suggesting a generally brighter or clearer spectral quality. However, in the lower SPL range (below 70 dB), its performance becomes less consistent, as shown by scattered red and green cells—indicating possible instability or noisiness in soft speech. For CPPs, the comparison shows UnivNet achieving predominantly higher values (indicated by the predominant green coloring), particularly in the louder SPL range (above 65 dB). This implies more stable harmonic structure and stronger voice quality when synthesizing louder or more effortful speech in UnivNet.

## 4. Discussion

This pilot study has demonstrated the effectiveness of voice mapping-based tools in objective TTS evaluation. It captures subtle voice quality differences that might be overlooked by traditional metrics and subjective tests. Utterance-level comparisons can be conducted using cycle-by-cycle metric extraction and statistics, while system-level assessments can be represented through voice mapping.

The voice range serves as the most intuitive metric for demonstrating a model's capability - the closer it matches the raw voice, the higher the flexibility and variability of the model's speech generation. In our tests, VITS demonstrated the largest voice range (measured in semitone × dB cells).

Mapping suggests that crest factor, the peakiness of the acoustic voice signal, is often an indicator for bad synthesis or artifacts, even though in most models the crest factor difference from the raw speech remains small. The spectrum balance proved valuable for assessing frequency balance, which contributes to speech intelligibility, especially in soft phonation regions. The CPPs is related to the voice periodicity. For the natural female speech in LJSpeech, the CPPs is typically in the range 7-8 dB. Within this range, higher values indicate better periodicity, though overly high values that exceed 10 dB might indicate robotic-sounding output.

The voice mapping visualization proves especially valuable in analyzing voice quality variations across different phonation modes, such as soft or loud speech. For instance, our ability to identify Glow-TTS's superior performance in soft phonation could inform the development of models better suited for reproducing natural quiet speech segments.

Ranking the voice quality of TTS models involves numerous influencing factors. For synthesized speech, a key metric is the deviation from the training data. Our method effectively complements CMOS, but as an objective, signal processing-based, and reproducible approach. It is capable of generalizing across various models.

While numerous synthesis approaches exist, including articulatory synthesis, formant synthesis, concatenative synthesis, and earlier neural methods (FNNs, LSTMs), a comprehensive voice mapping comparison across all these models presents significant challenges. We could hypothesize that concatenative synthesis might achieve voice quality similar to natural speech but show limited voice range variability. Similarly, articulatory synthesis might face challenges in both quality and range aspects. However, these remain speculative without empirical validation through voice mapping analysis. Such cross-model analysis requires using the same database for valid comparisons.

One limitation of the LJSpeech dataset and many other popular TTS datasets is the lack of sound pressure level (SPL) calibration, which complicates the application of voice mapping. Ideally, a calibrated dataset with real-world SPL values would allow for more accurate comparisons between synthetic and natural voices. However, we observed that the TTS models in discussion do not significantly modulate volume and apply normalization, which preserves the original speaker's intensity patterns. Although these intensity levels do not correspond to actual physical SPL values, they still offer perceptible loudness differences in synthesized speech. Another limitation lies in the dataset selection. This study relies solely on the LJSpeech corpus for both training and reference, while many modern TTS systems are trained on a mixture of datasets that include multiple speakers, languages, and emotions. Such diverse training data make it difficult to isolate the effect of the model architecture alone. Consequently, our current design that requires a tightly controlled conditions may seem less generalizable to today's multi-condition models. This limitation also implies that some recent and more powerful models cannot be evaluated under the "comparison with the original voice" paradigm. Extending this framework to more diverse datasets, multi-speaker scenarios and even to voice conversion assessment will be part of our future work.

In addition, we could integrate additional metrics such as jitter, shimmer, HNR into this evaluation framework. Importantly, these metrics should be analyzed on a cycle-by-cycle basis and mapped to their corresponding $f_o$ and SPL coordinates, as both fundamental frequency and sound pressure level substantially influence these measurements[61, 66]. We would also explore the potential of generating voice maps during model training, allowing for real-time comparison and iterative improvements.

## 5. Conclusion

We have proposed a voice quality comparison framework based on cycle-synchronized metric extraction and voice mapping visualization. Voice range serves as the primary indicator of model capability - the closer it matches the raw voice range, the better the model's ability to generate varied and flexible speech output.

Our analysis through multiple metrics revealed some quality indicators: crest factor identifies synthesis artifacts through lower values, spectrum balance effectively evaluates frequency distribution particularly in soft phonation, and CPPs reflects voice periodicity where values matching the raw speaker indicate better quality, while overly high values exceeding 10 suggests robotic output. This is probably due to a combination of excessive period regularity and the upper end of the voice spectrum being over-emphasized.

Among the evaluated models, VITS demonstrated strong overall performance in maintaining natural voice qualities, while Glow-TTS showed superior performance in soft phonation despite limited voice range. In vocoder comparisons, UnivNet outperformed Multiband-MelGAN in preserving both spectral characteristics and periodicity.

## Acknowledgments

This work was supported by the KTH-CSC programme (grant number 202006010113). The notion of using voice maps to assess TTS was suggested by Joakim Gustafson. The authors would also like to thank Gustav Eje Henter and Olov Engwall for their helpful discussions and feedback during the development of this work. The authors also acknowledge support from Språkbanken Tal, during the revision phase of this manuscript.

## Author Declarations

### *Conflict of Interest*

The authors declare that there is no conflict of interest.

### *Authors' contributions*

Author Sten Ternström conceptualized the study. Author Huanchen Cai designed and performed the experiments, developed the data processing pipeline and computational methods, processed the data, analyzed the results, and wrote the original draft. Author Sten Ternström contributed to methodology development, results interpretation, and manuscript review and editing.

*Ethics Approval*

This study utilized pre-existing speech datasets and did not involve direct experimentation with human participants or animals. The research concerned the computational analysis of publicly available voice recordings and the evaluation of open-source speech synthesis models.

**Data Availability**

The data and models used in this study are publicly available. The text-to-speech models were implemented using the Coqui toolkit (https://github.com/coqui-ai/TTS) and ESPnet toolkit (https://github.com/espnet/espnet). The speech synthesis experiments were conducted using the publicly available LJSpeech dataset (https://keithito.com/LJ-Speech-Dataset/).